\documentclass[letter]{aa}  

\usepackage{graphicx}
\usepackage{txfonts}
\usepackage{lipsum}
\usepackage{subcaption}         % necessary for continued figures, example in section 3
\usepackage{lscape}             % to rotate a single page table, example in appendix.
\usepackage{placeins}           % useful with \FloatBarrier, to keep 
\usepackage[]{hyperref}
\hypersetup{colorlinks=true,citecolor=blue}
\usepackage{multirow}
\usepackage{makecell}
\usepackage{multicol}
\usepackage{wrapfig}

\begin{document}

   \title{A comparative analysis of the presence of haze in sub-Neptunes observed with JWST}

   \titlerunning{haze in sub-Neptunes observed with JWST}
   % \subtitle{}

   \author{A. Y. Jaziri\inst{1,2},
        T. Drant\inst{3}
        \and
        M. Turbet\inst{4,2}
        }

   \authorrunning{Jaziri et al.}

   \institute{LATMOS/IPSL, UVSQ Universit\'{e} Paris-Saclay, Sorbonne Universit\'{e}, CNRS, Guyancourt, France
            \and Laboratoire d’astrophysique de Bordeaux, Univ. Bordeaux, CNRS, B18N, all\'{e}e Geoffroy Saint-Hilaire, 33615 Pessac, France
            \and 
            ETH University, Center for Origin and Prevalence of Life, Department of Earth and Planetary Sciences, 8092 Zurich, Switzerland.
            \and
            Laboratoire de M\'{e}t\'{e}orologie Dynamique/IPSL, CNRS, Sorbonne Universit\'{e}, Ecole Normale Sup\'{e}rieure, PSL Research University, Ecole Polytechnique, 75005 Paris, France\\
            \email{yassin.jaziri@latmos.ipsl.fr}}

   \date{Received ??; Accepted ??}

% \abstract{}{}{}{}{}
% 5 {} token are mandatory

  \abstract
   {Photochemical hazes have been proposed to explain the diversity of sub-Neptune transmission spectra, with early HST observations suggesting a temperature-dependent trend in aerosol opacity. However, JWST observations reveal a more heterogeneous picture. Here, we perform a consistent atmospheric retrieval analysis of eight moderately irradiated JWST-observed sub-Neptunes, LP~791-18~c, TOI-1231~b, TOI-732~c, TOI-270~d, GJ~1214~b, GJ~9827~d, GJ~3090~b, and TOI-421~b, using experimentally derived haze optical properties. We compare haze-free models with two experimentally haze analogues to assess haze evidence and spectral signatures. We find significant evidence for haze in LP~791-18~c, TOI-732~c and GJ~3090~b, while LP~791-18~c and GJ~3090~b evidences may be biased by transit light source effects (TLS). Haze-favored solutions are primarily driven by short-wavelength spectral slope rather than characteristic infrared absorption features of the experimental analogues, except for TOI-732~c. We find no clear correlation between retrieved CH$_4$ abundance and haze preference, suggesting that haze formation also depends on atmospheric composition and photochemical conditions. For TOI-270~d, the inferred haze properties vary with the MIRI inter-instrument offset and temperature parameterization without yielding significant evidence for haze, highlighting multidimensional retrieval degeneracies. Overall, our results do not support a ubiquitous haze trend across the JWST sub-Neptune sample, but rather a heterogeneous picture that could nevertheless evolve toward the HST trend with improved characterization of the host stars, more accurate TLS corrections and accounting for different haze types. Broader experimental haze measurements and multi-wavelength observations, including phase curves, will be essential to disentangle atmospheric aerosols from stellar contamination and other retrieval degeneracies.
}

   \keywords{exoplanets -- atmospheres –- hazes -- JWST}

   \maketitle
   \nolinenumbers

%%%%%%%%%%%%%%%%%%%%%%%%%%%%%%%%%%%%%%%%%%%%%%%%%%%%%%%%%%%%%%
\section{Introduction}

Sub-Neptunes are among the most abundant classes of exoplanets, yet have no counterpart in the Solar System. Their intermediate nature between terrestrial planets and gas giants makes them key laboratories for studying planetary formation, atmospheric evolution, and chemical diversity \citep{Bean2021}. Transmission spectroscopy has revealed a broad diversity of atmospheric spectra, ranging from prominent molecular features to nearly flat spectra \citep{Madhusudhan2025b}. A leading explanation for this diversity is the presence of aerosols \citep{Gao2021}. Clouds produced by equilibrium condensation and hazes generated by disequilibrium photochemistry can both obscure molecular features and modify the apparent atmospheric scale height, but arise from different physical processes and therefore provide distinct insights into atmospheric chemistry.

Early HST transmission spectra suggested population-level trends in aerosol opacity, with warm sub-Neptunes generally exhibiting more muted spectral features \citep{Yu2021, Estrela2022, Edwards2023, Brande2024}. These results motivated the view that haze formation may be common in warm hydrogen-rich atmospheres and may depend on parameters such as temperature, metallicity, and composition. JWST observations have since revealed a more heterogeneous picture \citep{Roy2026}, raising questions about the prevalence of hazes and whether the HST trends may partly reflect observational limitations, retrieval assumptions, or simplified aerosol models. Haze formation is expected to depend on a complex interplay between atmospheric composition, irradiation, redox conditions, metallicity, and vertical transport \citep{Yu2021, Lavvas2021, Gao2023, Drant2026, Lavvas2026, Li2026}. Recent laboratory experiments further indicate that aerosols can form over a broader range of atmospheric conditions than predicted by conventional hydrocarbon photochemistry and that their optical properties can differ substantially from commonly adopted parameterizations \citep{He2024, Drant2026}.

\cite{Jaziri2025} incorporated experimentally derived exoplanet haze optical properties from \cite{Drant2026} into atmospheric retrieval analyses. Applied to JWST observations of K2-18~b, this framework suggested that aerosol opacities may reconcile spectral behaviors that are difficult to reproduce with standard retrieval assumptions, and that aerosol signatures may be visible among molecular features. Their results indicate that experimentally informed aerosol modeling can substantially alter the interpretation of atmospheric composition.

Motivated by these results, we extend this approach to a homogeneous re-analysis of eight moderately irradiated sub-Neptunes observed with JWST: LP~791-18~c \citep{Roy2026}, TOI-1231~b \citep{Sarkar2026}, TOI-732~c \citep{Rigby2025}, TOI-270~d \citep{Holmberg2024, Constantinou2026, Holmberg2026}, GJ~1214~b \citep{Schlawin2024}, GJ~9827~d \citep{Piaulet-Ghorayeb2024}, GJ~3090~b \citep{Ahrer2025}, and TOI-421~b \citep{Davenport2025}. We restrict the sample to planets with JWST observations extending below 2.5~$\mu$m, where the wavelength-dependent haze slope can be constrained, except GJ~1214~b, for which independent studies provide evidence for haze \citep{Kempton2023, Ohno2025}. Our objectives are twofold: first, to test whether the aerosol signatures and spectral behavior identified for K2-18 b can be reproduced across a wider population of sub-Neptunes; and second, to investigate whether such a consistent treatment of aerosols can provide new insight into the physical mechanisms governing haze formation. By analyzing this sample within a common retrieval framework, we aim to reassess the observational evidence for haze trends in the JWST era and evaluate whether a unified picture of aerosol formation in sub-Neptune atmospheres can emerge. 

\section{Methods}
\label{sect:method}

We used transmission spectra from previous JWST analyses (see Table~\ref{tab:spectra}). Atmospheric retrievals were performed with \texttt{TauREx 3}\footnote{\url{https://github.com/ucl-exoplanets/taurex3}} \citep{Al-Refaie2021} and the \texttt{TauREx-PyMieScatt}\footnote{\url{https://github.com/groningen-exoatmospheres/taurex-pymiescatt}} module \citep{Changeat2025}, which accounts for photochemical hazes following \citet{Jaziri2025}. The module assumes spherical particles and computes their optical properties using Mie theory, with laboratory-derived refractive indices as input; fractal aggregates are not considered. Following \citet{Jaziri2025}, we restricted the analysis to the exo1 and exo2 haze analogues (Table~\ref{tab:haze}). Exo1 differs from the other analogues by lacking nitrogen-bearing functional groups and exhibiting only C-H$_x$ vibrational signatures. It was favored in the K2-18~b retrievals because its absorption features overlap with molecular features in the observed spectrum without introducing strong additional features inconsistent with the data. Exo2 is also nitrogen-free but has a more aromatic structure, resulting in stronger C=C-related absorption features, which were less consistent with the K2-18~b spectrum and made it less favored than exo1. We therefore use exo1 and exo2 as complementary haze tests.

\begin{table}[h]
 \centering
 \caption[]{\label{tab:haze}Summary of the different haze refractive index data considered from \cite{Drant2026}}
\begin{tabular}{lccc}
 \hline \hline
Tholins      & Gas composition  & Experimental setup    \\ \hline
\noalign{\smallskip}
exo1     & 95 \% Ar - 5\% CH$_4$ & PAMPRE            \\
exo2     & 95 \% Ar - 5\% CH$_4$ & COSmIC            \\
\hline
\end{tabular}
\end{table}

We performed a complementary analysis to the original observational studies, focusing specifically on assessing the significance of experimentally motivated haze models relative to a haze-free atmosphere. For each combined dataset, we therefore carried out three retrievals: a haze-free model, an exo1 haze model, and an exo2 haze model (see Appendix~\ref{ap:taurexret} for details). The retrievals included a broad set of molecular absorbers consistent with previous published analyses of these JWST datasets and with the molecules available in the ExoMol opacity database for hydrogen-rich atmospheres \citep{Chubb2021}: CH$_4$, CO$_2$, CO, H$_2$O, NH$_3$, C$_2$H$_2$, H$_2$S, SO$_2$, and OCS. We note that additional species, such as CS$_2$ and C$_2$H$_6$, could also be relevant, since they have been potentially detected in TOI-270~d or K2-18~b. However, we chose to remain consistent with using only ExoMol database while also limiting the substantial computational cost associated with retrievals including experimentally derived haze optical properties.

%%%%%%%%%%%%%%%%%%%%%%%%%%%%%%%%%%%%%%%%%%%%%%%%%%%%%%%%%%%%%%

\section{Statistical evidence for photochemical hazes}
\label{sec:haze_detection}

\begin{figure}
    \centering
    \includegraphics[width=1.\linewidth]{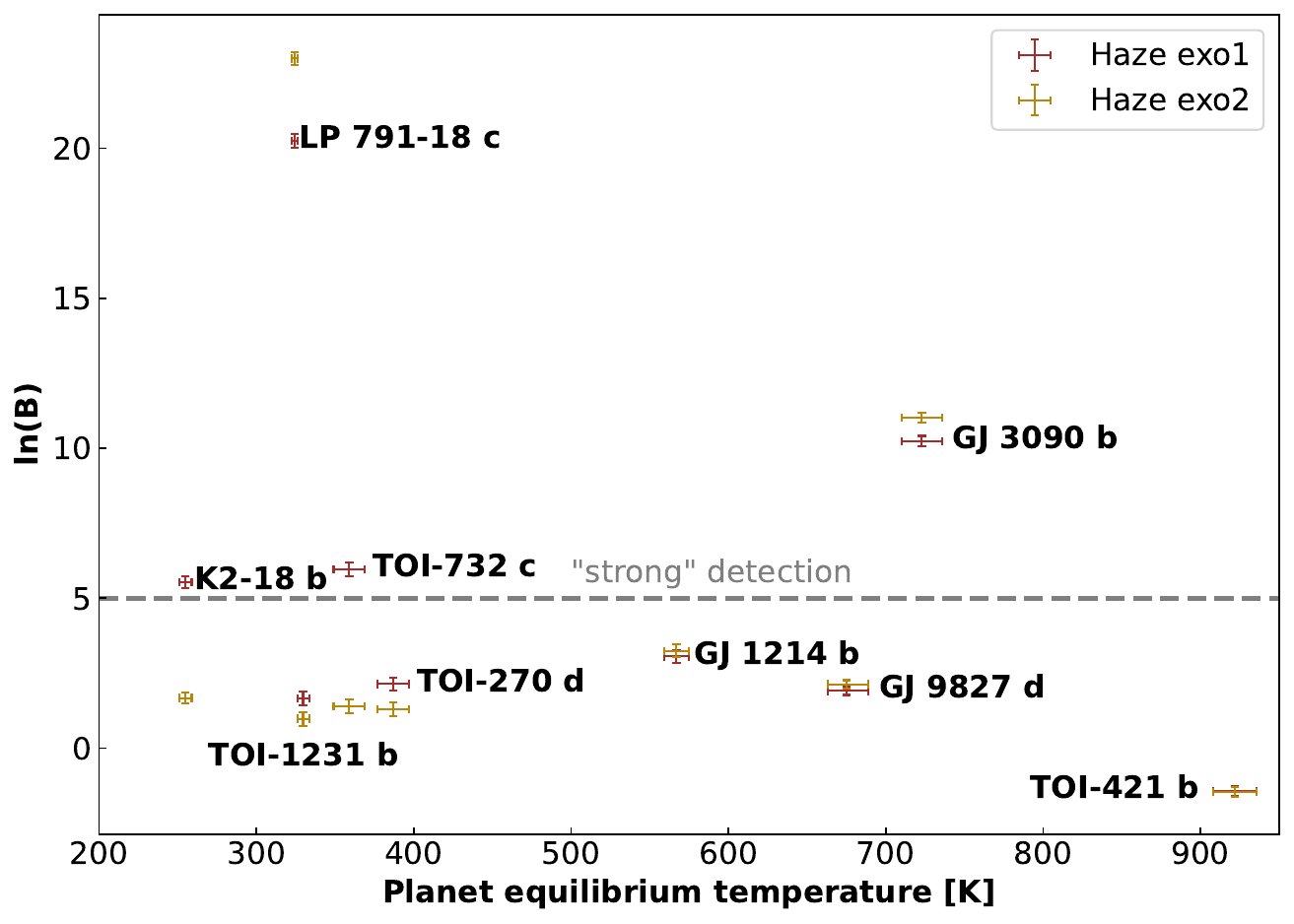}
    \includegraphics[width=1.\linewidth]{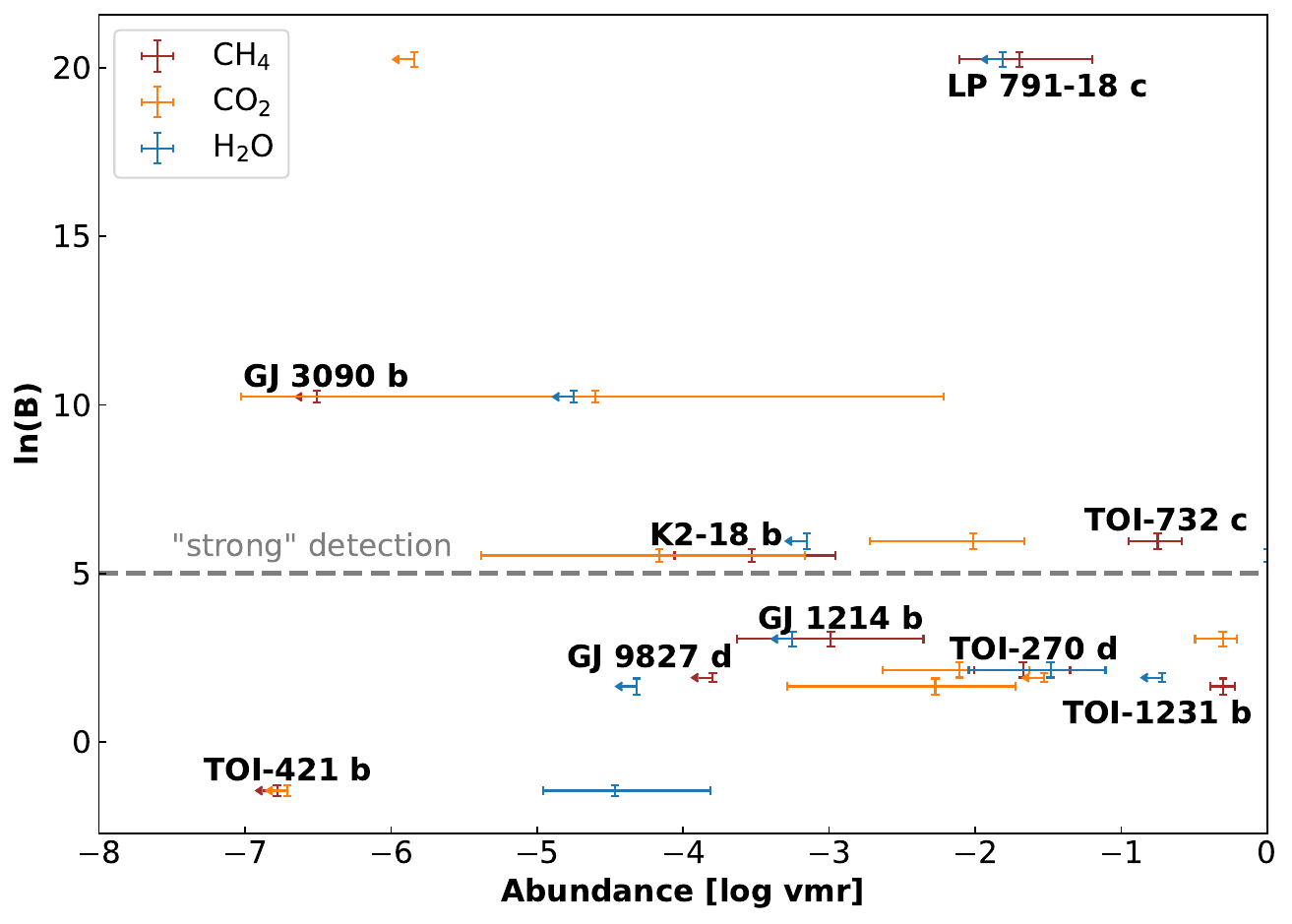}
    \caption{Bayes Factor versus equilibrium temperature (top) for the exo1 (brown) and exo2 (gold) haze models, and versus retrieved CH$_4$, CO$_2$, and H$_2$O abundances (bottom) for exo1 only. The strong-detection threshold, ln(B) = 5 \citep{Benneke2013}, is shown by the horizontal gray dashed line. Bayes Factors indicate model preference given the data, fitted parameters, and priors, not a definitive detection or absence of hazes. Only TOI-270~d results with a 0 ppm MIRI offset are shown.}
    \label{fig:bayes_teq_x}
\end{figure}

The Bayesian evidence obtained for the haze-free, exo1, and exo2 retrievals is summarized in Table~\ref{tab:retrievals}, while the corresponding Bayes factors, $\ln(B)$ \citep{Benneke2013}, are shown as a function of planetary and atmospheric parameters in Figs.~\ref{fig:bayes_teq_x},~\ref{fig:bayes_derived} and~\ref{fig:bayes_haze}. We find no ubiquitous preference for the experimentally motivated haze models: significant evidence for haze is found for only a subset of planets, while several targets are equally well or better described by haze-free atmospheres. No trend with equilibrium temperature is apparent (Fig.~\ref{fig:bayes_teq_x}).

The strongest haze evidence is found for LP~791-18~c, where both exo1 and exo2 are significantly preferred over the haze-free retrieval, consistent with the CH$_4$ detection and strong haze layer reported by \citet{Roy2026}. Although \citet{Roy2026} found limited transit light-source effect (TLSe), recent observations suggest that their synthetic stellar spectra may not accurately reproduce the host star below 1~$\mu$m \citep{Murray2026}. We also find significant haze evidence for GJ~3090~b. However, the NIRISS spectrum is likely affected by substantial TLSe \citep{Ahrer2025}. The haze preference for this target was obtained using the visit showing the most favorable spectral slope, making this detection particularly sensitive to the adopted dataset. The haze detections for both planets are therefore uncertain. Finally, TOI-732~c resembles K2-18~b with a haze exo1 detection but not exo2.

No significant preference for either haze model is found for TOI-1231~b, GJ~1214~b, GJ~9827~d, or TOI-421~b. For GJ~1214~b, this contrasts with previous evidence for clouds or hazes \citep{Kempton2023,Ohno2025}, although our retrieval, consistent with \citet{Ohno2025}, favors a high-metallicity, CO$_2$-dominated atmosphere (Fig.~\ref{fig:bayes_teq_x}) above the haze layer, reflecting the haze–metallicity degeneracy. GJ~9827~d similarly allows substantial aerosol opacity without requiring hazes, while TOI-421~b shows no improvement from haze models, consistent with \citet{Davenport2025}. For TOI-270~d, neither haze model is statistically favored, regardless of the MIRI offset (Appendices~\ref{ap:toi270d} and~\ref{ap:toi_nomiri}). Overall, Accounting for potentially biased detections in LP~791-18~c and GJ~3090~b and the dependence on haze type, Fig.~\ref{fig:bayes_teq_x} tentatively suggests a parabolic trend similar to that observed with HST, although the statistical evidence remains inconclusive.

These results highlight the distinction between retrieving a haze solution and finding statistical evidence for haze. Allowing aerosols and temperature variations (Appendix~\ref{ap:haze_degeneracies}) can provide additional flexibility to reproduce the observed spectra and yield constrained haze parameters, but does not necessarily mean that aerosols are required by the data. In particular, retrievals favoring a haze layer generally converge toward higher temperatures than their haze-free counterparts (Fig.~\ref{fig:temperature}), suggesting a degeneracy to influence the atmospheric scale height. The following sections therefore examine which spectral regions drive the inferred haze solutions and how they compete with alternative atmospheric scenarios.

%%%%%%%%%%%%%%%%%%%%%%%%%%%%%%%%%%%%%%%%%%%%%%%%%%%%%%%%%%%%%%

\section{Spectral signatures and properties of the retrieved hazes}
\label{sec:haze_signature}

The spectra corresponding to the different retrieval solutions are shown in Figs.~\ref{fig:haze1},~\ref{fig:haze2} and~\ref{fig:haze3}, with a summary of all planet with haze exo1 retrieved spectra Fig.~\ref{fig:specall}. A common feature of the haze-favored solutions is that the inferred aerosol opacity primarily affects the short-wavelength part of the transmission spectrum. In contrast, the characteristic infrared absorption features associated with the laboratory haze analogues \citep{Drant2026} generally have a smaller contribution to the overall fit.

For LP~791-18~c, both haze models provide a strong statistical preference over the haze-free case, with physically coherent retrieved haze properties (Fig.~\ref{fig:bayes_haze}). The haze is located at high altitude, with a mid pressure of approximately 1~Pa. Only CH$_4$ is constrained with the model without haze, but for the exo2 model, the haze parameters are constrained together with CH$_4$ and H$_2$S, while NH$_3$ is additionally constrained in the exo1 case. The haze solutions also imply a higher atmospheric mean molecular weight (Fig.~\ref{fig:bayes_derived}), increasing from 6.85 in the haze-free solution to 11.7 for exo1, but this is balanced with an increase of 200 K of the temperature in order to increase the scale height. Thus, the haze retrievals compensate for the additional high-altitude aerosol opacity through changes in the atmospheric scale height and composition. Despite the strong statistical preference for the haze models, no prominent infrared haze features are retrieved; instead, the haze signature is dominated by enhanced short-wavelength opacity, which is biased by TLSe.

This behavior is different to that found for K2-18~b by \citet{Jaziri2025}. In that case, the experimentally derived optical properties provided additional flexibility to reproduce the short-wavelength spectral shape while also allowing possible contributions from the infrared absorption bands of the haze analogues. The comparison with LP~791-18~c therefore suggests that the spectral slope may be a more important observable for identifying these haze solutions than the direct detection of their characteristic infrared vibrational features. On the contrary, TOI-732~c have same spectral behavior (Fig.~\ref{fig:haze3}) and retrieved atmospheric properties (Fig.~\ref{ap:bayes_derived} and~\ref{ap:bayes_haze}) as K2-18~b, which question again the TLSe on LP~791-18~c spectra biasing the haze retrieval.

For GJ~3090~b, the haze-favored solution is also driven by the short-wavelength spectral slope (Fig.~\ref{fig:haze2}), with an optically thick, high-altitude haze layer, $\sim$0.1~Pa (Fig.~\ref{fig:bayes_haze}), that flattens the transmission spectrum and mutes molecular features (Fig.~\ref{fig:bayes_teq_x}). Without haze, CO$_2$ is constrained at high abundance, providing a high-metallicity solution capable of reducing the atmospheric scale height and muting molecular features. When haze is included, CO and high-altitude aerosols partly take over this role. Neither solution shows prominent infrared molecular or haze features, consistent with the muted spectrum reported by \citet{Ahrer2025}. The apparent haze detection is therefore driven by the broadband spectral slope, which is biased by the strong TLSe in the NIRISS data.

The remaining targets illustrate that a muted transmission spectrum does not necessarily indicate the presence of photochemical haze. For GJ~1214~b, neither haze model yields constrained aerosol solution (Fig.~\ref{fig:bayes_haze}), and the spectrum is instead reproduced by a high-metallicity CO$_2$- and CH$_4$-rich atmosphere (Fig.~\ref{fig:bayes_teq_x}). Nevertheless, the phase curve analysis from \cite{Kempton2023} has revealed an albedo of the planet that favor the presence of cloud or haze. Distinguishing these scenarios requires better constraints on the short-wavelength slope, which may differ substantially between high-metallicity, cloud-dominated, and haze-dominated atmospheres. The existing HST/WFC3 data \citep{Kreidberg2014} cannot resolve this ambiguity, as an unknown offset between the HST and JWST spectra affects the inferred slope.

GJ~9827~d also illustrates the degeneracy between molecular composition and aerosol opacity. In the haze-free retrieval, H$_2$O is constrained, consistent with the results of \citet{Piaulet-Ghorayeb2024}, while SO$_2$ is also retrieved at high abundance. The latter increase the atmospheric mean molecular weight and reduce the spectral amplitude without producing prominent NIRISS features. 
Including haze weakens these constraints, as aerosols can similarly mute the spectrum. The exo1 model additionally suggests a haze contribution alongside a CO$_2$ feature in NIRSpec G395H, unlike exo2, but neither model is statistically preferred. Broader-wavelength observations are therefore needed to distinguish these scenarios. Clouds also remain a possible explanation for the muted spectrum, although water clouds are disfavored \citep{Turbet2023}.

Interestingly, TOI-270~d provides a useful comparison with K2-18~b, TOI-732~c and LP~791-18~c, which have similar retrieved CH$_4$ abundances but for which haze is more strongly favored (Fig.~\ref{fig:bayes_teq_x}). A high CH$_4$ abundance may provide an important source of carbon for photochemical haze production, but the efficiency of haze formation also depends on the atmospheric chemical environment. In particular, a more oxidized atmosphere with a high CO$_2$ abundance may inhibit or alter the pathways leading to the formation of complex hydrocarbon aerosols \citep{Trainer2004}. LP~791-18~c has a relatively low retrieved CO$_2$ abundance, whereas K2-18~b and TOI-732~c exhibits comparable CH$_4$ and CO$_2$ abundances to TOI-270~d while still showing evidence for haze. The absence of a clear correlation between the retrieved CH$_4$ abundance (or C/O ratio, see Fig.~\ref{fig:bayes_derived}) and the statistical preference for haze therefore suggests that the atmospheric abundance of CH$_4$ alone is insufficient to explain the diversity of haze signatures observed in our sample. Instead, haze production is likely governed by a combination of atmospheric composition, redox state, and photochemical pathways, with the stellar UV irradiation potentially playing an additional role. Investigating the relationship between stellar UV irradiation and the retrieved haze properties will therefore be an important avenue for future work, but additional UV measurement are needed for LP~791-18~c, TOI-270~d, GJ~3090~b, TOI-421~b. These trends should nevertheless be further investigated by accounting for additional degeneracies arising from stellar contamination.

%%%%%%%%%%%%%%%%%%%%%%%%%%%%%%%%%%%%%%%%%%%%%%%%%%%%%%%%%%%%%%
\section{Conclusions}
\label{sec:conclusion}

We presented a homogeneous re-analysis of eight JWST-observed sub-Neptunes using experimentally derived haze optical properties from \citet{Drant2026}, following the retrieval framework developed by \citet{Jaziri2025} for K2-18~b. By comparing haze-free, exo1, and exo2 retrievals, we investigated whether the haze solutions identified for K2-18~b are representative of a broader population and could explain the haze trend suggested by earlier HST observations.

Our results do not support a ubiquitous presence of photochemical hazes across the JWST sub-Neptune sample. Evidences for haze are found for LP~791-18~c, TOI-732~c and GJ~3090~b, while LP~791-18~c and GJ~3090~b evidences are likely biased by TLSe. No significant evidence for haze is found for TOI-1231~b, GJ~1214~b, GJ~9827~d, TOI-421~b, or TOI-270~d, for which muted spectra can instead be explained by high metallicity, molecular composition, or other aerosol scenarios such as clouds. When haze is favored, the retrievals are primarily driven by the broadband short-wavelength slope rather than by the infrared absorption features of the experimental haze analogues, except TOI-732~c which behave same as K2-18~b. Since the short-wavelength slope can also be produced by TLSe, the infrared features provide an important complementary diagnostic. Their absence in several haze-favored cases highlights the need to explore a broader range of experimentally derived haze optical properties and test in more detail their degeneracy with TLSe.

We find no clear correlation between the retrieved CH$_4$ abundance and the statistical preference for haze, suggesting that haze production depends on a combination of atmospheric composition and photochemical conditions rather than on CH$_4$ abundance alone. The TOI-270~d analysis further shows that variations in the MIRI offset can substantially modify the inferred haze properties without producing statistically significant evidence for haze. Together with the $\sim$7~$\mu$m haze feature retrieved for K2-18~b and TOI-732~c, these results highlight the importance of MIRI data for constraining hazes and motivate further re-analysis with the upcoming JWST MIRI observations of LP~791-18~c, GJ~3090~b, and GJ~9827~d.

Overall, our results suggest that the haze trend inferred from early HST observations does not imply a universal haze-forming mechanism among sub-Neptunes. Instead, JWST observations point toward a heterogeneous picture in which haze signatures depend on atmospheric composition and photochemical conditions. However this conclusion remains sensitive to stellar contamination and aerosols opacity and we cannot rule out the parabolic HST trend. Further joint retrievals that simultaneously account for haze and TLSe, future broad-wavelength observations, improved constraints on stellar heterogeneity, and laboratory measurements spanning a wider range of atmospheric compositions and haze formation conditions will be essential to determine whether a common mechanism governs haze formation across the sub-Neptune population.

%%%%%%%%%%%%%%%%%%%%%%%%%%%%%%%%%%%%%%%%%%%%%%%%%%%%%%%%%%%%%%

% \paragraph{\textit{\scriptsize Data availability}} {\scriptsize Retrieval data: \texttt{Zenodo} \href{https://doi.org/10.5281/zenodo.22282544}{doi:10.5281/zenodo.22282544}}

%%%%%%%%%%%%%%%%%%%%%%%%%%%%%%%%%%%%%%%%%%%%%%%%%%%%%%%%%%%%%%
\begin{acknowledgements}
This project has received funding from the European Research Council (ERC) under the ERC OxyPlanets projects (grant agreement No. 101053033). This project acknowledges funding from the European Research Council (ERC) under the European Union's Horizon 2020 research and innovation programme (grant agreement No. 679030/WHIPLASH). T.D. acknowledges support by the 2024 NOMIS-ETH postdoc fellowship program. T.D. thanks the NOMIS foundation and ETH Zurich for funding via the 2024 Fellowship program. This work was supported by CNES, focused on the NIRISS, NIRSpec and MIRI instruments on JWST. Some of the data presented in this paper were obtained from the Mikulski Archive for Space Telescopes (MAST) at the Space Telescope Science Institute. The specific observations analyzed can be accessed via \href{https://doi.org/10.17909/c8z7-r590}{doi:10.17909/c8z7-r590}, \href{https://doi.org/10.17909/1hrx-cn31}{doi:10.17909/1hrx-cn31}, \href{https://doi.org/10.17909/tqa7-th94}{doi:10.17909/tqa7-th94}, \href{https://doi.org/10.17909/sx19-8852}{doi:10.17909/sx19-8852}, \href{https://doi.org/10.17909/dvqh-2r48}{doi:10.17909/dvqh-2r48}, \href{https://doi.org/10.17909/r2bz-wb41}{doi:10.17909/r2bz-wb41}, \href{https://doi.org/10.17909/91wj-dt23}{doi:10.17909/91wj-dt23}, \href{https://doi.org/10.17909/3ds1-8z15}{doi:10.17909/3ds1-8z15} and \href{https://doi.org/10.17909/rx29-yw62}{doi:10.17909/rx29-yw62}.
\end{acknowledgements}

\bibliographystyle{aa}
\bibliography{biblio}

%%%%%%%%%%%%%%%%%%%%%%%%%%%%%%%%%%%%%%%%%%%%%%%%%%%%%%%%%%%%%%%
% Appendices must be placed after   \end{thebibliography}
% They will be placed automatically on a new page.
%%%%%%%%%%%%%%%%%%%%%%%%%%%%%%%%%%%%%%%%%%%%%%%%%%%%%%%%%%%%%%%
\begin{appendix}

\onecolumn

%%%%%%%%%%%%%%%%%%%%%%%%%%%%%%%%%%%%%%%%%%%%%%%%%%%%%%%%%%%%%%

\section{Retrieval setup}
\label{ap:taurexret}

\begin{wraptable}{R}{0.4\linewidth}
\centering
\caption{Free parameters and priors for the retrievals.}
\label{tab:priors}
\begin{tabular}{c|c}
    \hline
    \textbf{Parameters} & \textbf{Bounds} \\ \hline
    \multicolumn{2}{c}{\textbf{Common}} \\ \hline
    T$_{top}$ [K]                 & 50  to 1500   \\
    T$_{bot}$ [K]                 & 300 to 2000  \\
    T$_{1}$ [K]                   & 50  to 1500   \\ 
    T$_{2}$ [K]                   & 50  to 1500   \\
    log$_{10}$(P$_{1}$) [Pa]      & 2   to 6     \\
    log$_{10}$(P$_{2}$) [Pa]      & -2  to 5     \\
    radius [R$_{jup}$]            & 0.1 to 0.3   \\
    \hline
    \multicolumn{2}{c}{\textbf{Chemistry}} \\ \hline
    log$_{10}$(H$_2$O)            & -12 to -0.15 \\
    log$_{10}$(CO)                & -12 to -0.15 \\
    log$_{10}$(CH$_4$)            & -12 to -0.15 \\
    log$_{10}$(CO$_2$)            & -12 to -0.15 \\
    log$_{10}$(OCS)               & -12 to -0.15 \\
    log$_{10}$(NH$_3$)            & -12 to -0.15 \\
    log$_{10}$(C$_2$H$_2$)        & -12 to -0.15 \\
    log$_{10}$(SO$_2$)            & -12 to -0.15 \\
    log$_{10}$(H$_2$S)            & -12 to -0.15 \\
    \hline
    \multicolumn{2}{c}{\textbf{Aerosols}} \\ \hline
    log($\mu$) [$\mu$m]              & -2 to 1  \\
    log($\chi$) [molecules.m$^{-3}$] & 4  to 13 \\
    log(P) [Pa]                      & -2 to 5  \\
    log($\Delta$P) [Pa]              & 0  to 5  \\
\end{tabular}
\end{wraptable}

Our analysis is limited to free retrievals, in which the atmospheric composition is parameterized by vertically constant volume mixing ratios for each chemical species. This approach does not impose chemical equilibrium or self-consistent chemistry, allowing the molecular abundances to be varied independently rather than being constrained by an assumed chemical network. We note, however, that for molecular species with weak or undetected spectral features, the retrieved abundances can remain strongly prior-dependent.

For all retrievals, the planetary radius was treated as a free parameter using the same prior range for every target. Given the relatively similar radii of the planets in our sample, a common prior encompasses all values while avoiding the introduction of target-dependent assumptions. The atmospheric temperature profile was parameterized using the 4-point profile implemented in \texttt{TauREx 3}. We adopted this parameterization because it represents the minimum level of complexity required to allow for thermal inversions and assess the influence of aerosols on the thermal structure of the upper atmosphere. The adopted temperature priors are listed in Table~\ref{tab:priors}. Each retrieval was performed using 750 live points in the multinest sampling algorithm.

The retrievals included vertically constant abundances for all molecular species listed in Table~\ref{tab:priors}. These molecules correspond to the species most commonly investigated in previous studies of hydrogen-rich sub-Neptune atmospheres and for which high-quality opacity data are available in the ExoMol database \citep{Chubb2021}. The priors adopted for both the molecular abundances and the aerosol parameters are summarized in Table~\ref{tab:priors}. N$_2$ was not included as a retrieved species, trying to reduce the parameter space and computational time, because it does not produce prominent molecular absorption features over the wavelength range considered here. However, N$_2$ would affect the transmission spectrum indirectly through the mean molecular weight. This effect is partially degenerate with other parameters already free in our retrievals, in particular the reference radius, temperature profile, and bulk H$_2$/He composition. The omission of N$_2$ could therefore affect the inferred absolute atmospheric composition and mean molecular weight, but we do not expect strong biases in the haze detection and characterization.

We did not include the transit light-source effect (TLSe) as a free component in the retrievals. Including TLSe as an additional free component would substantially increase the dimensionality of the retrievals and introduce an additional degeneracy with the atmospheric radius, temperature structure, and aerosol properties, particularly at short wavelengths. A systematic joint retrieval of TLSe and haze for the entire sample would therefore require a substantial extension of the present computational framework. We instead discuss the potential impact of stellar contamination on the haze interpretation for the individual targets where it is expected to be most relevant, and emphasize that haze constraints based on short-wavelength slopes should be interpreted with caution. From their respective analyses, the transmission spectra analyzed here that are sensitive of more than 100ppm TLSe amplitude are the one of LP~791-18~c and GJ~3090~b, while others are found less than 30ppm TLSe amplitude. The TLSe amplitudes of LP~791-18~c and GJ~3090~b have the same order of magnitude of the apparent slop at lower wavelength which bias the haze retrievals. While \citet{Roy2026} found TLSe of $\sim$50ppm for LP~791-18~c going to a 5.36$\sigma$ to a 3.54$\sigma$ haze detection including TLSe, \citet{Murray2026} suggest that their synthetic stellar spectra may not accurately reproduce the host star below 1 $\mu$m which would increased the TLSe that we suggest to be more than 100ppm.

The adopted planetary and stellar parameters are listed in Table~\ref{tab:parameters}, while the JWST datasets used in this work are summarized in Table~\ref{tab:spectra}. To assess the presence of haze we used the larger spectral coverage from NIRSOSS to MIRI (0.6 to 15 $\mu$m) whenever possible. We did not include the HST/WFC3 observations of GJ~1214~b, which are commonly combined with the JWST data to constrain the presence of clouds, because the unknown offset between the HST and JWST spectra could bias our investigation of haze signatures. For datasets combining multiple JWST instruments, we adopted the inter-instrument offsets reported in the original analyses. The only exception is TOI-270~d, for which the recently published MIRI observations have not yet been analysed jointly with the NIRISS and NIRSpec data. For this target, we therefore explored three possible MIRI offsets (0 and $\pm50$ ppm) to assess the sensitivity of our conclusions to the unknown absolute offset.

\paragraph{\textit{Data availability}} {Retrieval data: \texttt{Zenodo} \href{https://doi.org/10.5281/zenodo.22282544}{doi:10.5281/zenodo.22282544}}

\begin{table}[!htbp]
\centering
\caption{\label{tab:parameters} Summary of the planetary and stellar parameters adopted in this work.}
\begin{tabular}{lcccccc}
\hline\hline
Planet & R$_p$ [R$_{\rm Jup}$] & M$_p$ [M$_{\rm Jup}$] & a [AU] & R$_s$ [R$_{\odot}$] & T$_s$ [K] & Reference \\
\hline
LP 791-18 c & 0.220  & 0.022   & 0.0315  & 0.182  & 2960. & \cite{Roy2026} \\
TOI-1231 b  & 0.326  & 0.0485  & 0.1288  & 0.476  & 3553. & \cite{Burt2021} \\
TOI-732 c   & 0.213  & 0.0253  & 0.0757  & 0.380  & 3358. & \cite{Bonfanti2024} \\
TOI-270 d   & 0.1903 & 0.0150  & 0.07210 & 0.378  & 3506. & \cite{VanEylen2021} \\
GJ 1214 b   & 0.2438 & 0.0265  & 0.01505 & 0.2162 & 3101. & \cite{Mahajan2024} \\
GJ 9827 d   & 0.177  & 0.00950 & 0.053   & 0.58   & 4236. & \cite{Piaulet-Ghorayeb2024} \\
GJ 3090 b   & 0.194  & 0.0142  & 0.0316  & 0.516  & 3707. & \cite{Lamontagne2026} \\
TOI-421 b   & 0.236  & 0.021   & 0.0554  & 0.866  & 5291. & \cite{Krenn2024} \\
\hline
\end{tabular}
\end{table}

\begin{table}[!htbp]
\centering
\caption{\label{tab:spectra} Summary of transmission spectra adopted in this work.}
\begin{tabular}{lcc}
\hline\hline
Planet & JWST instruments (offset) & Reference \\
\hline
LP 791-18 c & NIRSpec PRISM & \cite{Roy2026} \\
TOI-1231 b  & NIRISS SOSS + NIRSpec G395H (+79.38ppm) & \cite{Sarkar2026} \\
TOI-732 c   & NIRISS SOSS (-112ppm) + NIRSpec G395H + MIRI (-112ppm) & \cite{Rigby2025} \\
% TOI-270 d   & NIRISS SOSS (+32ppm) + NIRSpec G395H + MIRI (0,$\pm$50ppm) & \cite{Holmberg2024, Constantinou2026, Holmberg2026} \\
TOI-270 d   & NIRISS SOSS (+32 ppm) + NIRSpec G395H + MIRI (0, $\pm$50 ppm) & \makecell[c]{\cite{Holmberg2024};\\ \cite{Constantinou2026};\\ \cite{Holmberg2026}} \\
GJ 1214 b   & NIRSpec G395H + MIRI (+50ppm) & \cite{Schlawin2024} \\
GJ 9827 d   & NIRISS SOSS & \cite{Piaulet-Ghorayeb2024} \\
GJ 3090 b   & NIRISS SOSS \texttt{exoTEDRF} + NIRSpec G395H \texttt{exoTEDRF!} & \cite{Ahrer2025} \\
TOI-421 b   & NIRISS SOSS \texttt{exoTEDRF} + NIRSpec G395H (+20ppm) \texttt{Eureka!} & \cite{Davenport2025} \\
\hline
\end{tabular}
\end{table}

\newpage

%%%%%%%%%%%%%%%%%%%%%%%%%%%%%%%%%%%%%%%%%%%%%%%%%%%%%%%%%%%%%%

\section{Summary of retrieval evidence and transmission spectra}
\label{ap:summary}

\begin{table}[h]
 \centering
 \caption[]{\label{tab:retrievals}Retrieval results for all observations}
\begin{tabular}{l|cc|cc|cc|cc}
 \hline \hline
Model                    & ln(Z)    & ln(B)     & ln(Z)    & ln(B)     & ln(Z)    & ln(B)     & ln(Z)    & ln(B)     \\ \hline
% \noalign{\smallskip}
 & \multicolumn{2}{c|}{LP~791-18~c} & \multicolumn{2}{c|}{TOI-270~d +50ppm for MIRI} & \multicolumn{2}{c|}{GJ~1214~b} & \multicolumn{2}{c}{TOI-421~b}\\ \hline
% \noalign{\smallskip}
Haze-free                & 2389.71  & Reference & 1296.02  & Reference & 594.76   & Reference & 587.22   & Reference \\
exo1                     & 2409.96  & 20.25     & 1298.25  & 2.23      & 597.82   & 3.06      & 585.78   & -1.44     \\
exo2                     & 2412.71  & 23.00     & 1297.86  & 1.84      & 598.00   & 3.24      & 585.75   & -1.47     \\ \hline
% \noalign{\smallskip}
 & \multicolumn{2}{c|}{TOI-1231~b} & \multicolumn{2}{c|}{TOI-270~d} & \multicolumn{2}{c|}{GJ~9827~d} \\ \hline
% \noalign{\smallskip}
Haze-free                & 13041.07 & Reference & 1301.91  & Reference & 1300.04  & Reference \\
exo1                     & 13042.72 & 1.65      & 1304.04  & 2.13      & 1301.94  & 1.90      \\
exo2                     & 13042.05 & 0.98      & 1303.20  & 1.29      & 1302.16  & 2.12      \\ \hline
% \noalign{\smallskip}
 & \multicolumn{2}{c|}{TOI-732~c} & \multicolumn{2}{c|}{TOI-270~d -50ppm for MIRI} & \multicolumn{2}{c|}{GJ~3090~b} \\ \hline
% \noalign{\smallskip}
Haze-free                & 17996.81 & Reference & 1296.59  & Reference & 1016.87  & Reference \\
exo1                     & 18002.77 & 5.96      & 1299.44  & 2.85      & 1027.10  & 10.23     \\
exo2                     & 17998.20 & 1.39      & 1299.01  & 2.42      & 1027.89  & 11.02     \\ \hline
\hline
\end{tabular}
\end{table}

\begin{figure}[!htbp]
    \centering
    \includegraphics[width=0.85\linewidth]{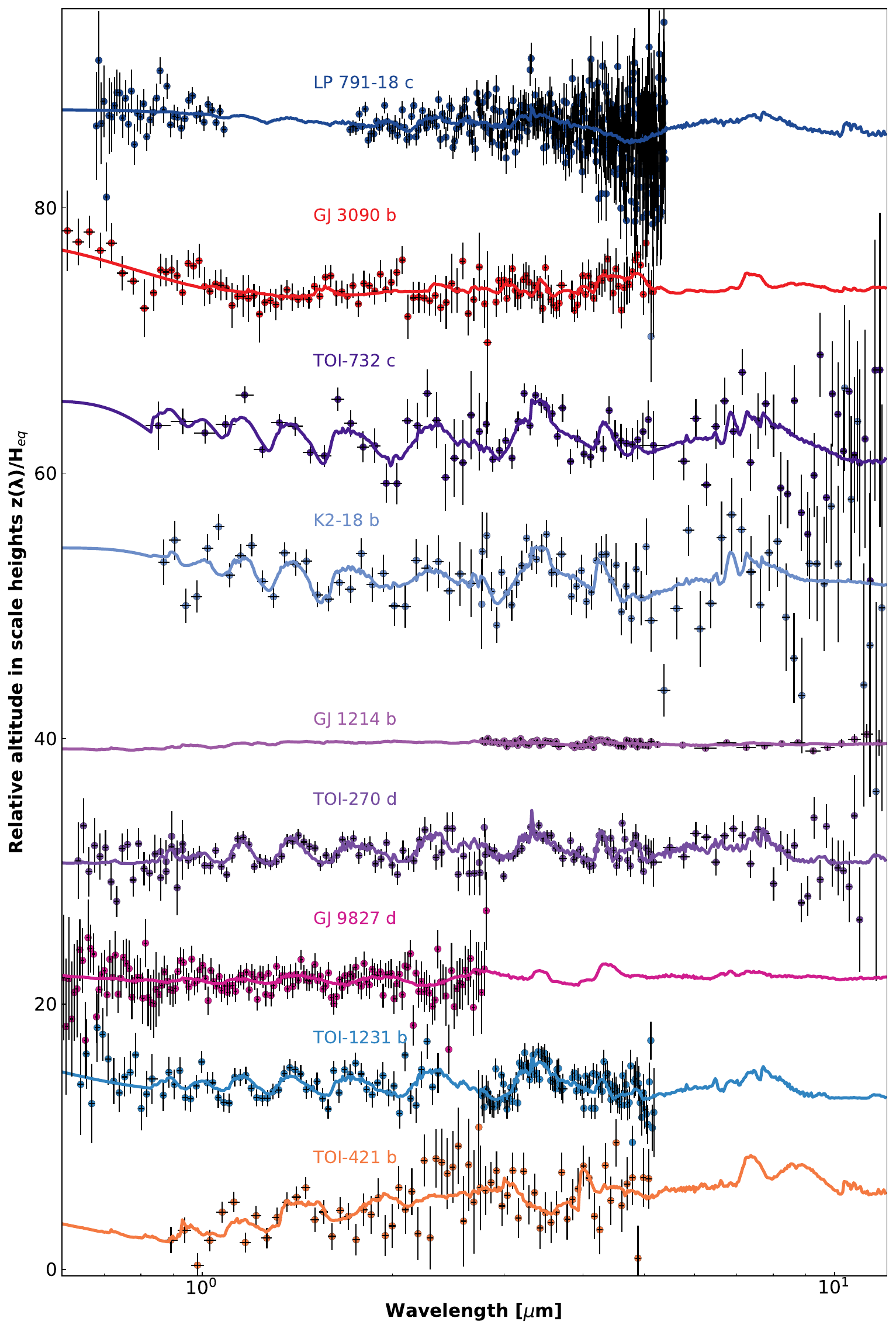}
    \caption{Observed spectra, as listed in Table~\ref{tab:spectra}, using the MIRI 0~ppm offset spectrum for TOI-270~d, together with the K2-18~b results from \citet{Jaziri2025}. The retrieved spectra including the exo1 haze are shown for each planet.}
    \label{fig:specall}
\end{figure}

Fig.~\ref{fig:specall} expresses spectra in units of the atmospheric scale height, $H_{\rm eq}=kT_{\rm eq}/(\mu g)$, estimated using the planet-specific equilibrium temperature and assuming a H/He atmosphere with a mean molecular weight of $\mu=2.3$~amu, following \citet{Sing2016}. The spectra are colored from blue to orange according to the equilibrium temperature of the planets, from lowest to highest. For clarity, the spectra are vertically shifted according to their $\ln(B)$ values (see Table~\ref{tab:retrievals}), with increasing vertical position corresponding to increasing statistical evidence for the presence of haze. Shifted values expressed in relative altitude are: K2-18~b:35 , LP~791-18~c:82, TOI-1231~b:10, TOI-732~c:51, TOI-270~d:26, GJ~1214~b:38, GJ~9827~d:19, GJ~3090~b:72, and TOI-421~b:0.

\clearpage

%%%%%%%%%%%%%%%%%%%%%%%%%%%%%%%%%%%%%%%%%%%%%%%%%%%%%%%%%%%%%%

\section{Bayes Factor as a function of the derived parameters}
\label{ap:bayes_derived}

\begin{figure}[!htbp]
    \centering
    \includegraphics[width=0.49\linewidth]{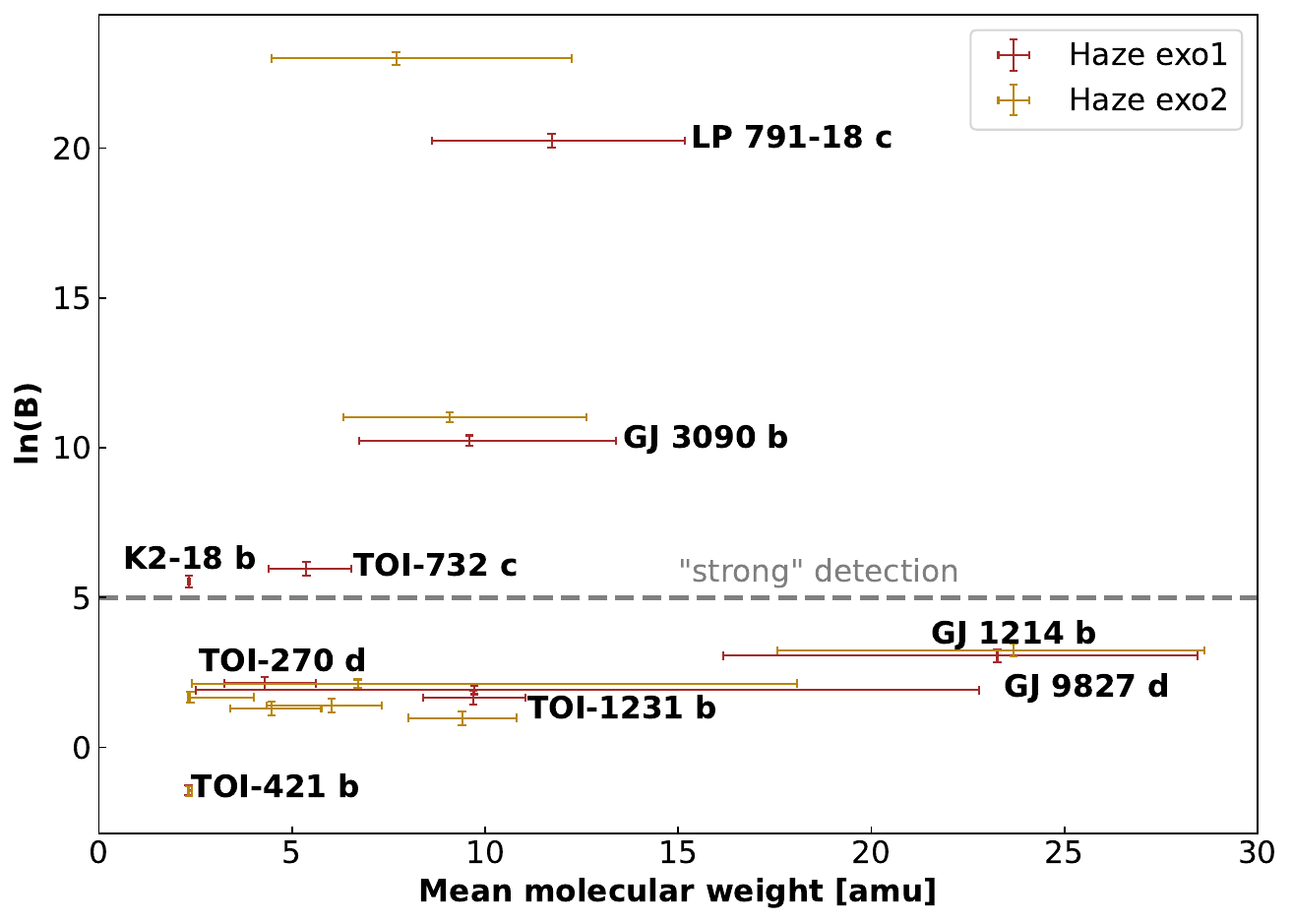}
    \includegraphics[width=0.49\linewidth]{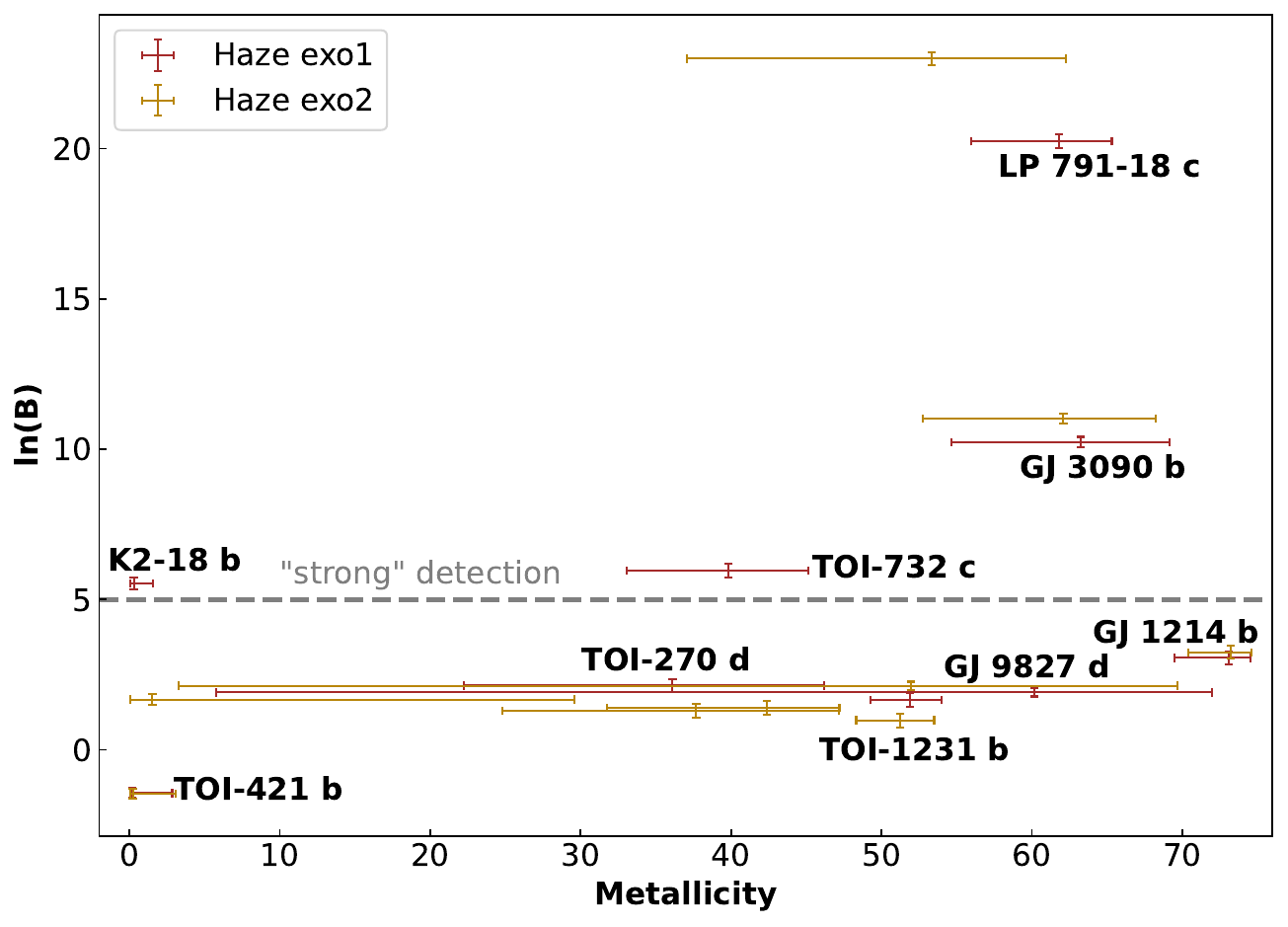}
    \includegraphics[width=0.49\linewidth]{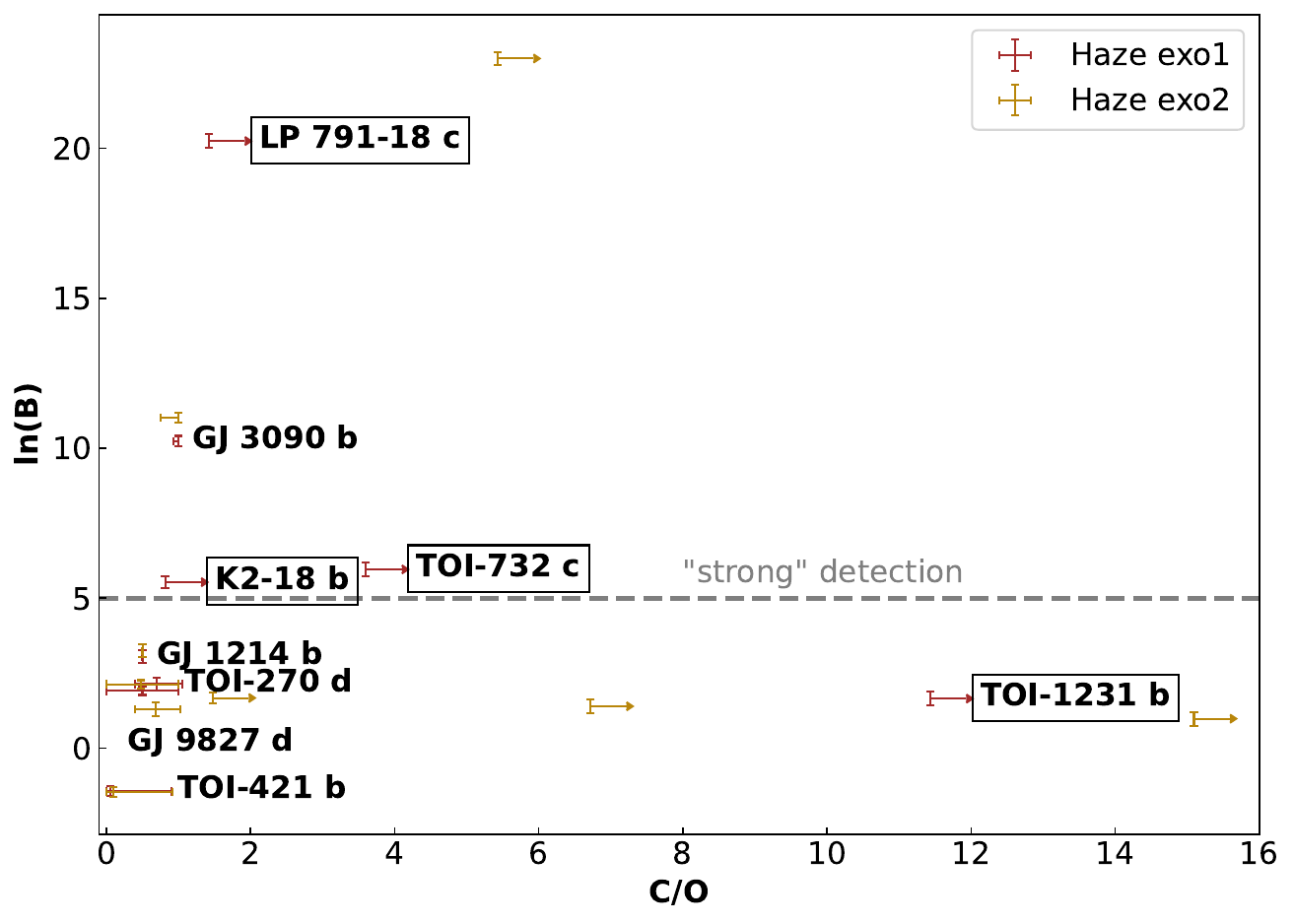}
    \caption{Bayes Factor as a function of the derived parameters for both exo1 (brown) and exo2 (gold) hazes. The limit of a strong detection at ln(B) = 5 \citep{Benneke2013} is highlighted with a horizontal dash gray line. This is not a final conclusion of the presence or the absence of hazes, but a preference of model considering the data, the fitted parameters and their priors. When an upper or lower limit is retrieved it is marked by a arrow. Large errorbars approximately centered means that the parameter is not constrained. Only TOI-270~d results with 0 offset of the MIRI observation are shown. C/O ratio above or potentially above 1 are highlighted by square boxes. Such high ratio is found due to high CH$_4$ retrieved abundance with no or poor constrain on oxidized molecules, such as CO$_2$ and H$_2$O, which biased these values. Better constrained on oxidized molecules would be needed to infer actual C/O ratio and link it to haze presence.}
    \label{fig:bayes_derived}
\end{figure}

%%%%%%%%%%%%%%%%%%%%%%%%%%%%%%%%%%%%%%%%%%%%%%%%%%%%%%%%%%%%%%

\section{Haze, temperature interplay}
\label{ap:haze_degeneracies}

\begin{figure}[!htbp]
    \centering
    \includegraphics[width=0.9\linewidth]{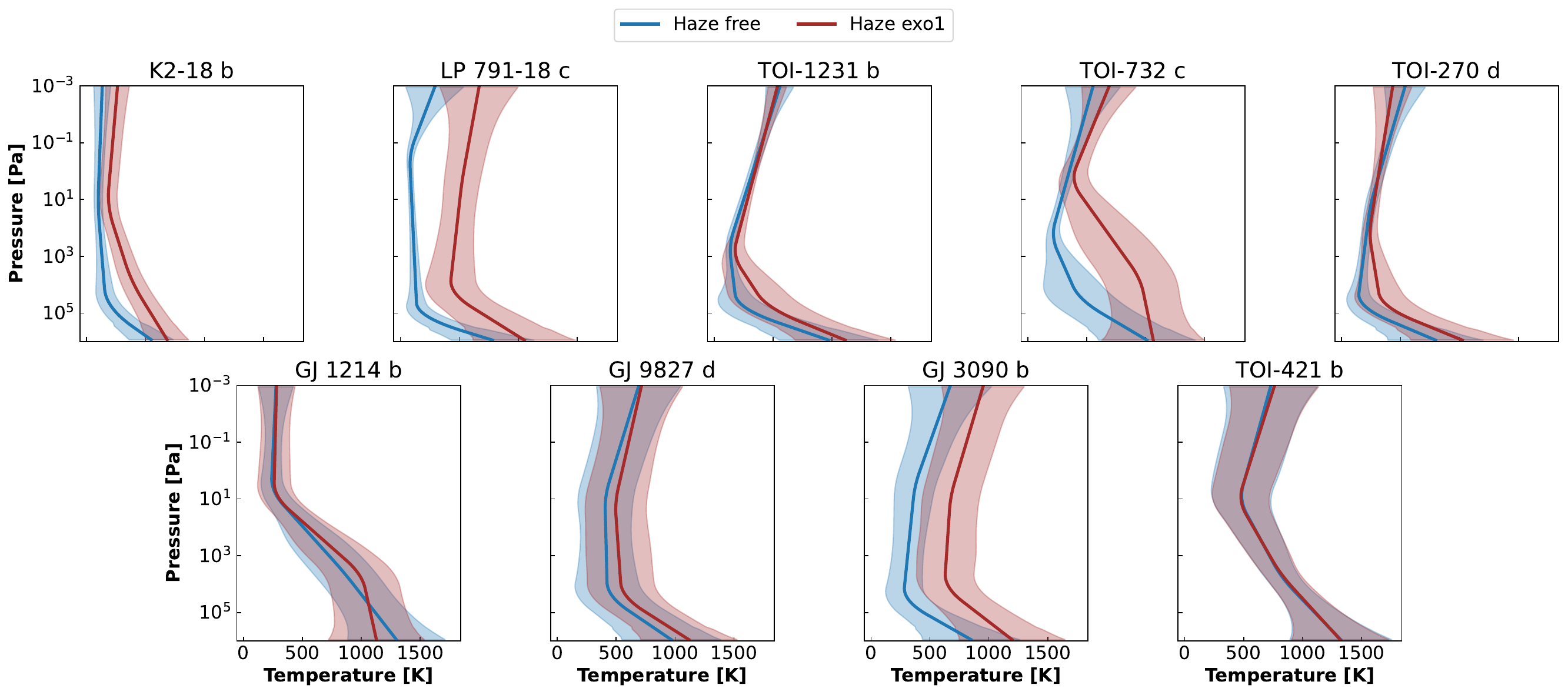}
    \caption{Temperature profiles retrieved for each planets. The blue curves correspond to the haze-free model, while the brown curves correspond to the haze exo1 model. Shaded regions indicate the 1$\sigma$ uncertainties on the retrieved temperature profiles. The comparison illustrates the impact of including haze opacity on the inferred atmospheric temperature structure.}
    \label{fig:temperature}
\end{figure}

The retrievals that favor a haze layer generally converge toward higher temperatures than their corresponding haze-free solutions, with the temperature increasing toward the upper atmosphere (Fig.~\ref{fig:temperature}). This behavior is qualitatively consistent with the results obtained for K2-18~b by \citet{Jaziri2025}.

This behavior can be understood in terms of the atmospheric scale height. A higher temperature increases the atmospheric scale height, allowing molecular absorption features to remain visible even when an optically significant haze is present at high altitude. The haze can then suppress the spectral amplitude at shorter wavelengths while the increased scale height maintains molecular features at longer wavelengths. The retrieval can therefore reproduce the observed spectrum through a combination of high-altitude haze and a relatively extended atmosphere.

These results highlight an intrinsic limitation of interpreting aerosol properties from transmission spectra alone. A reduced spectral amplitude can result from a combination of high atmospheric metallicity, clouds, photochemical hazes, or changes in the temperature structure. Consequently, a retrieval that allows additional aerosol parameters can find plausible haze solutions even when the data do not provide statistically significant evidence for haze.

\newpage
% \clearpage

%%%%%%%%%%%%%%%%%%%%%%%%%%%%%%%%%%%%%%%%%%%%%%%%%%%%%%%%%%%%%%

\section{Bayes Factor as a function of the haze retrieved parameters}
\label{ap:bayes_haze}

\begin{figure}[!htbp]
    \centering
    \includegraphics[width=0.45\linewidth]{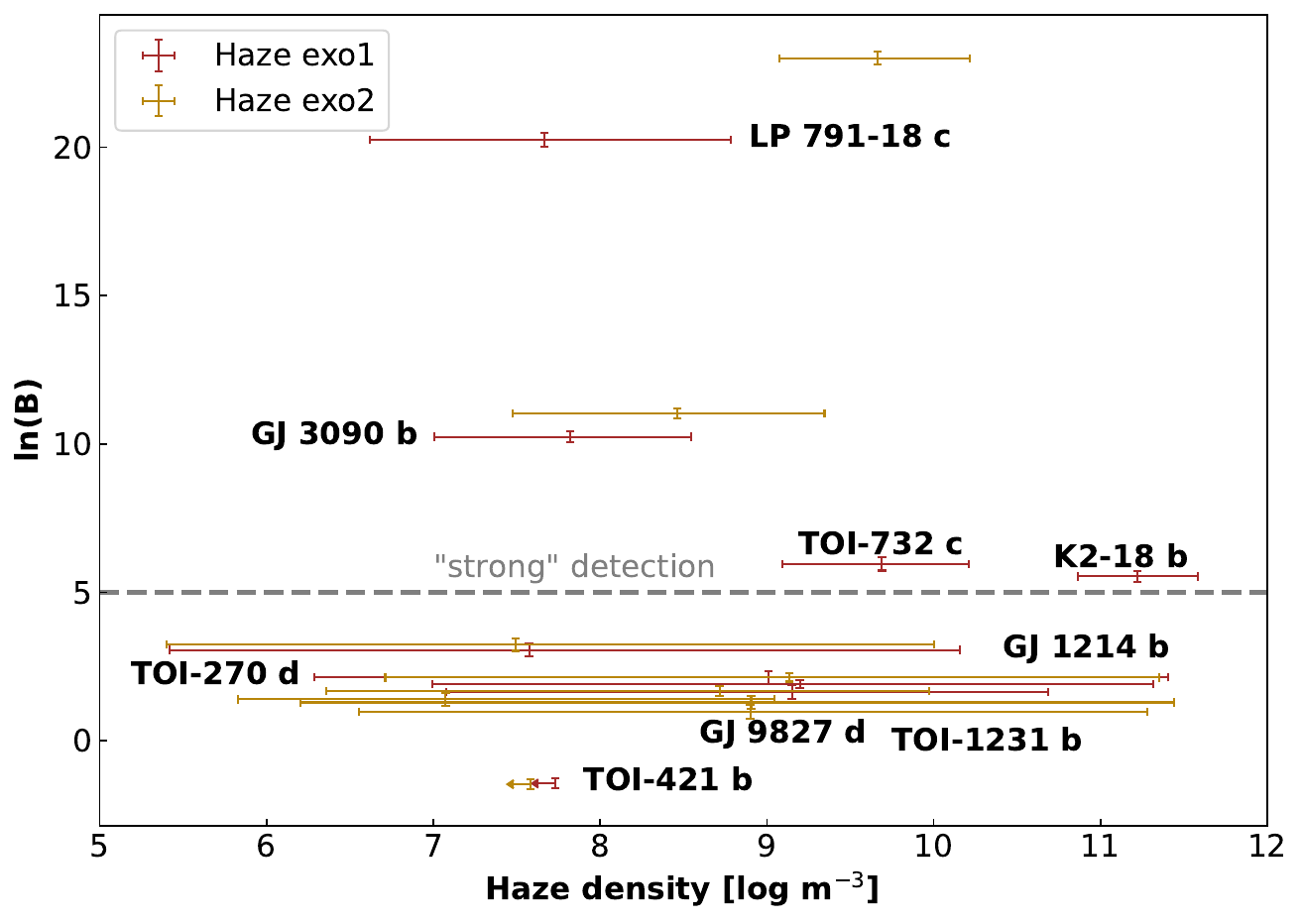}
    \includegraphics[width=0.45\linewidth]{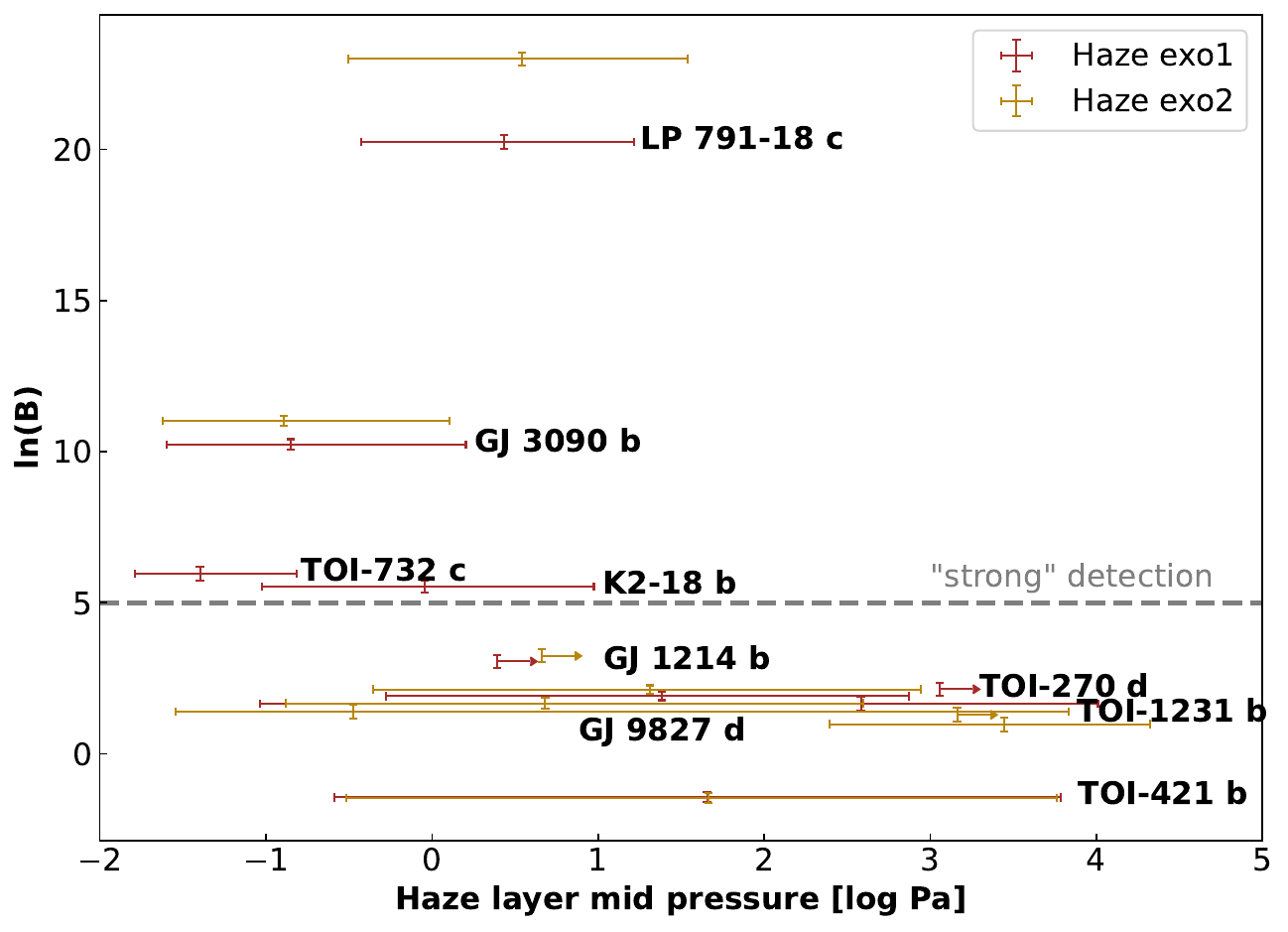}
    \includegraphics[width=0.45\linewidth]{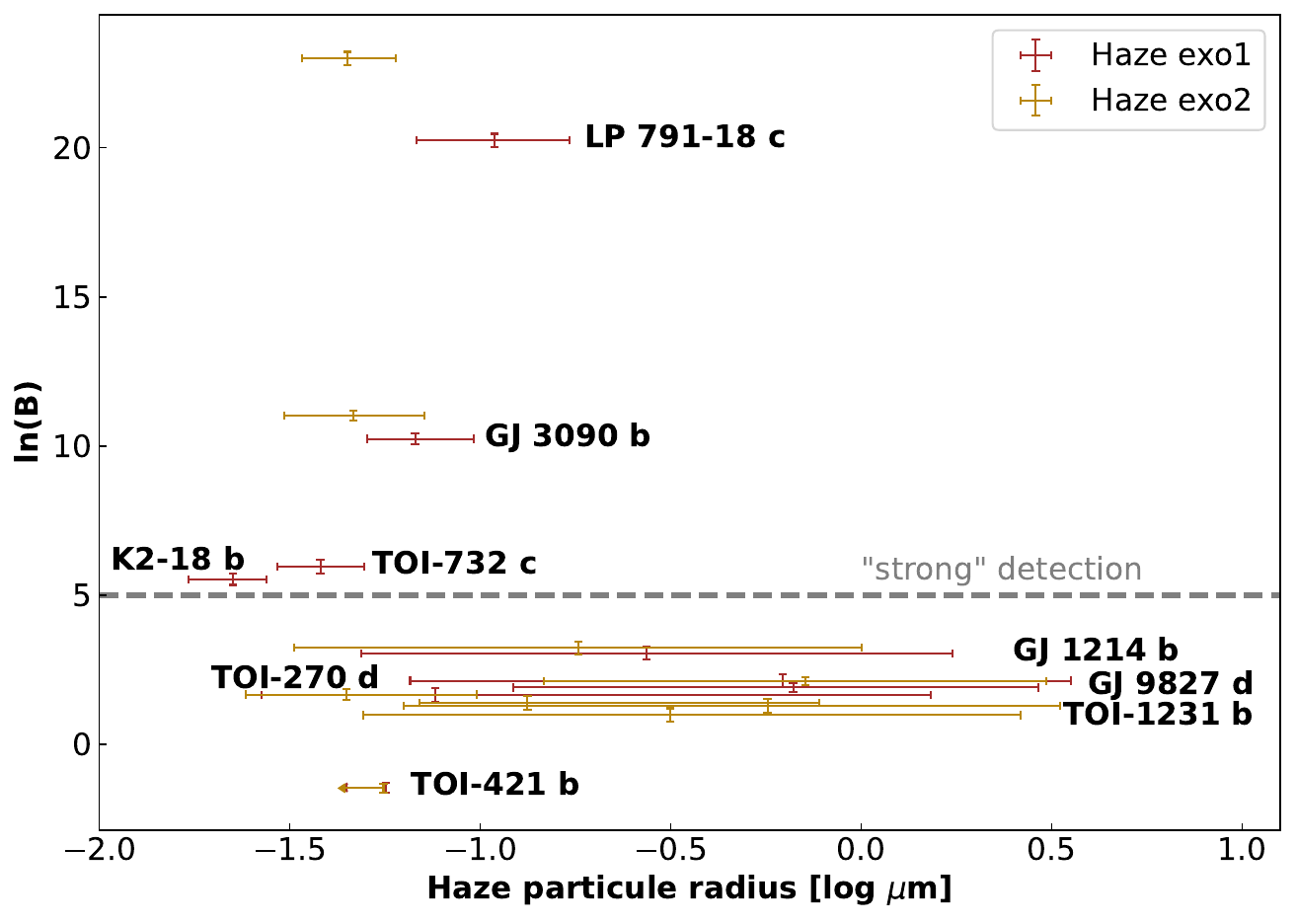}
    \caption{Bayes Factor as a function of the haze retrieved parameters for both exo1 (brown) and exo2 (gold) hazes. The limit of a strong detection at ln(B) = 5 \citep{Benneke2013} is highlighted with a horizontal dash gray line. This is not a final conclusion of the presence or the absence of hazes, but a preference of model considering the data, the fitted parameters and their priors. When an upper or lower limit is retrieved it is marked by a arrow. Large errorbars approximately centered means that the parameter is not constrained. Only TOI-270~d results with 0 offset of the MIRI observation are shown.}
    \label{fig:bayes_haze}
\end{figure}

\newpage

%%%%%%%%%%%%%%%%%%%%%%%%%%%%%%%%%%%%%%%%%%%%%%%%%%%%%%%%%%%%%%

\section{Retrieved planet spectra with haze contributions}
\label{ap:spec_contrib}

\begin{figure}[!htbp]
\centering
\includegraphics[width=0.98\linewidth]{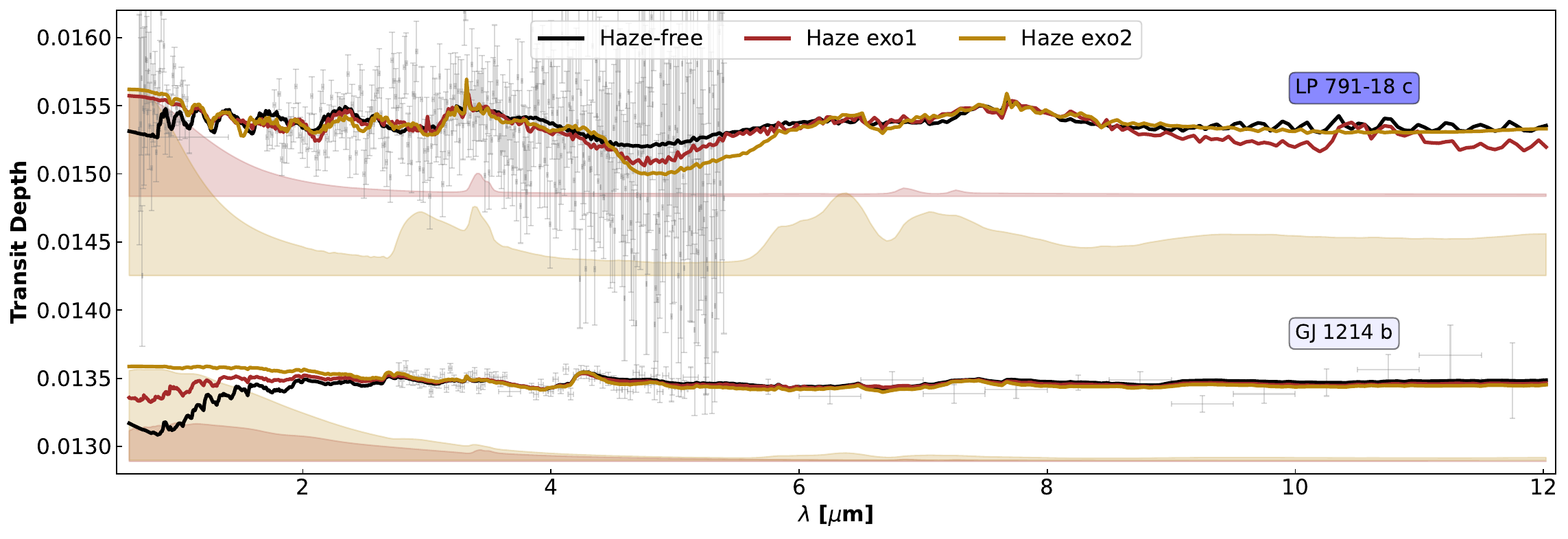}
  \caption{Observed spectra of LP~791-18~c and GJ~1214~b shown alongside are the best-fitting models (solid lines) for haze-free (black), haze exo1 (brown) and haze exo2 (yellow) retrievals, as well as their haze contribution.}
     \label{fig:haze1}
\end{figure}

\begin{figure}[!htbp]
\centering
\includegraphics[width=0.98\linewidth]{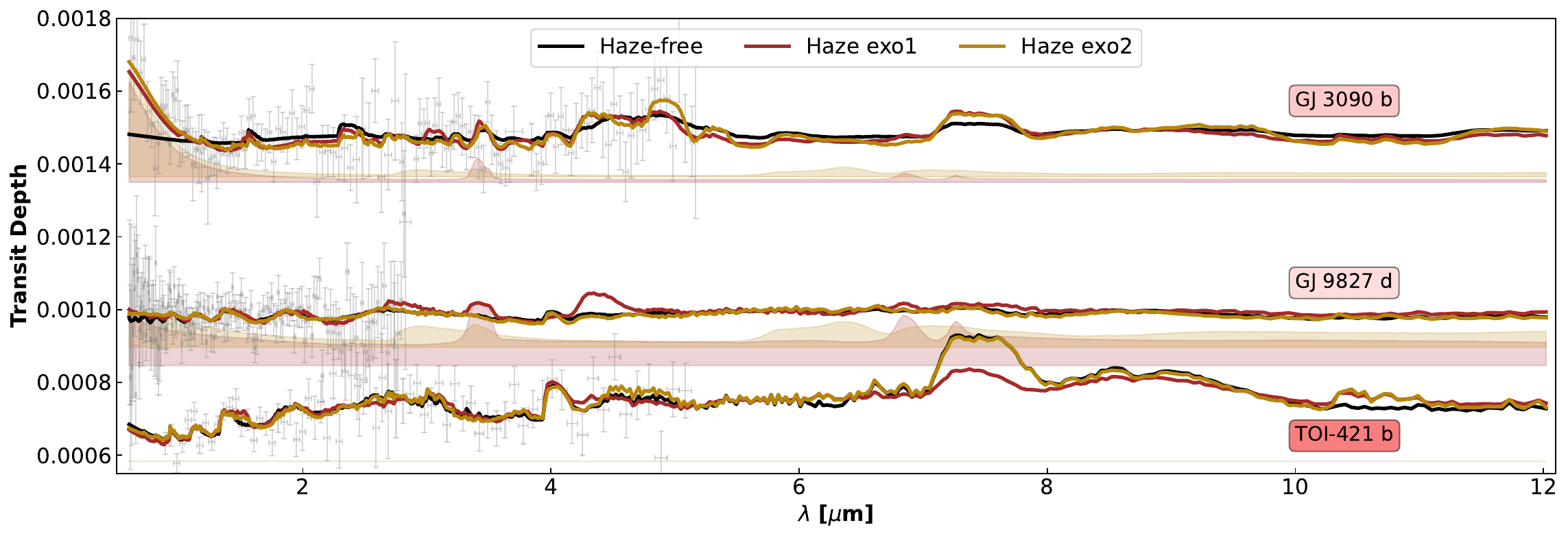}
  \caption{Observed spectra of GJ~3090~b, GJ~9827~d and TOI-421~b shown alongside are the best-fitting models (solid lines) for haze-free (black), haze exo1 (brown) and haze exo2 (yellow) retrievals, as well as their haze contribution.}
     \label{fig:haze2}
\end{figure}

\begin{figure}[!htbp]
\centering
\includegraphics[width=0.98\linewidth]{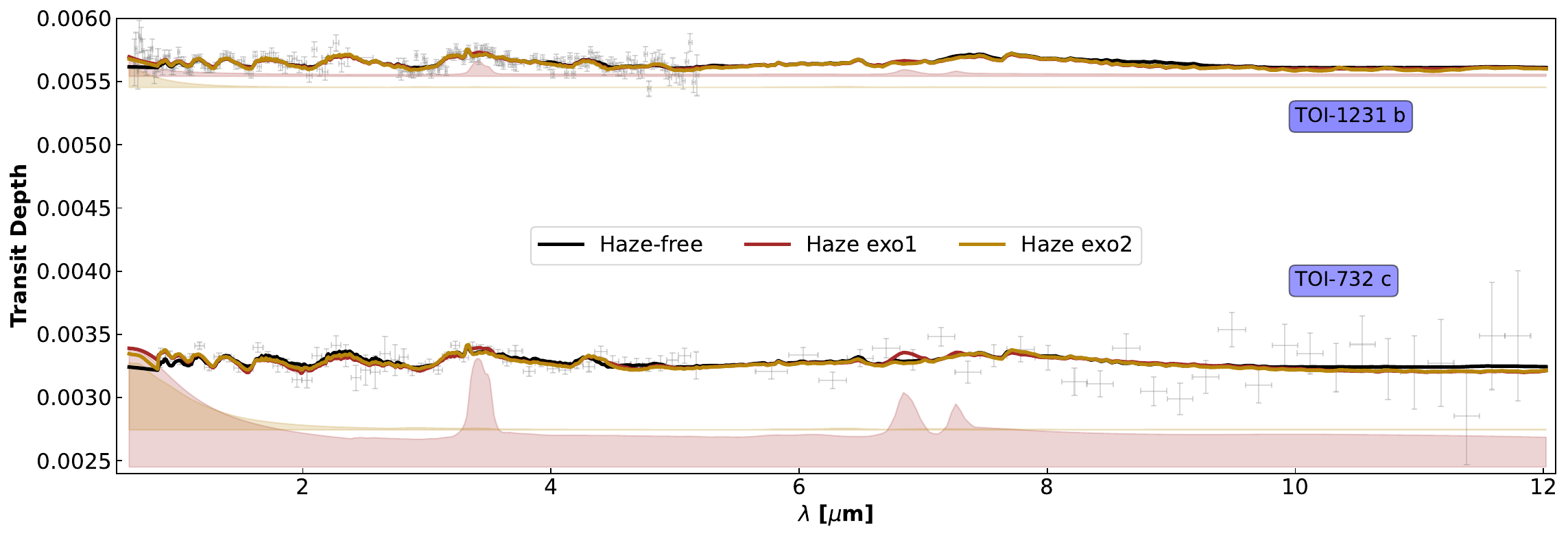}
  \caption{Observed spectra of TOI-1231~b and TOI-732~c shown alongside are the best-fitting models (solid lines) for haze-free (black), haze exo1 (brown) and haze exo2 (yellow) retrievals, as well as their haze contribution.}
     \label{fig:haze3}
\end{figure}

\section{TOI-270~d analysis without MIRI observation}
\label{ap:toi_nomiri}

Additional retrievals were performed for TOI-270~d using the same setup described in Appendix~\ref{ap:taurexret}. We considered the combined NIRISS SOSS and NIRSpec G395H spectra from \citet{Benneke2024}, at their native spectral resolution and without applying any inter-instrument offset.

\begin{figure*}[!htbp]
\centering
\includegraphics[width=\hsize]{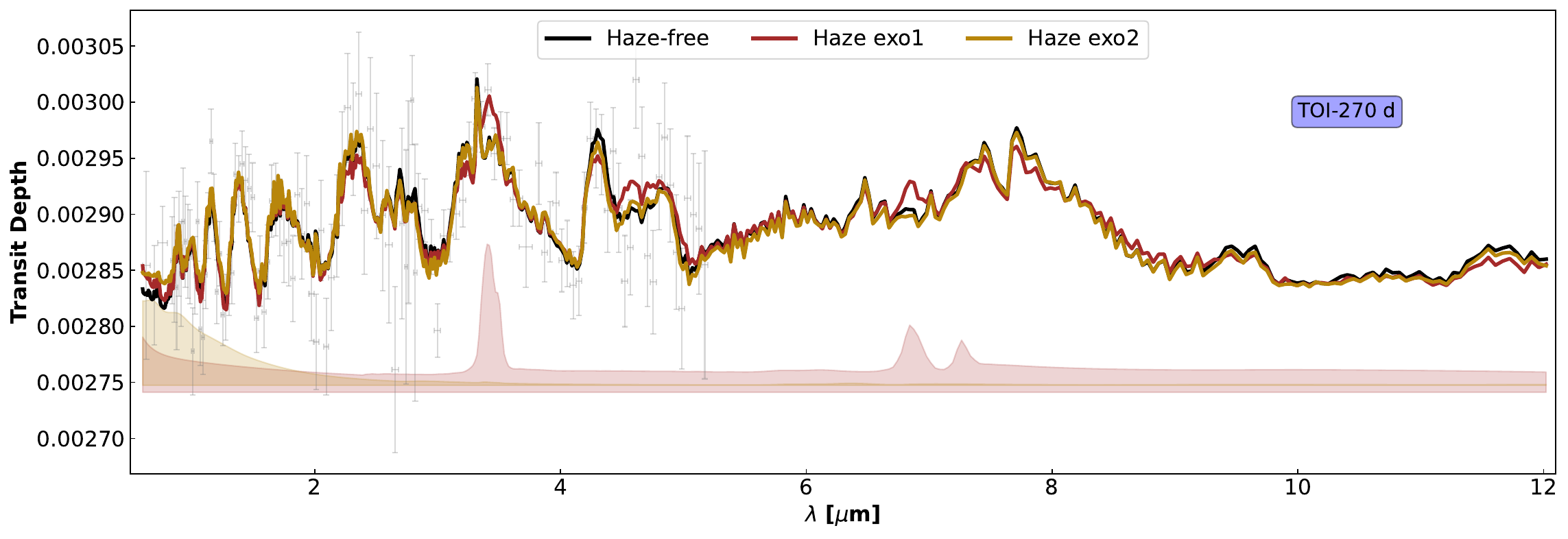}
  \caption{Combined spectra of NIRISS SOSS/NIRSpec G395H low-resolution data \citep{Benneke2024}. Shown alongside are the best-fitting models (solid lines) for haze-free (black), haze exo1 (brown) and haze exo2 (yellow) retrievals, as well as their haze contribution.}
     \label{fig:spec_toi_nomiri}
\end{figure*}

Compared to our retrievals including the MIRI data, the retrievals based only on NIRISS SOSS and NIRSpec G395H result in lower extended atmosphere and more tightly constrained haze parameters for the exo1 case. The retrieved haze is located at relatively high altitudes, and the corresponding haze opacity has a significant impact on the transmission spectrum, particularly around the 3.5~$\mu$m feature (Fig.~\ref{fig:spec_toi_nomiri}). However, the presence of the haze is not statistically favored over the haze-free model, with a Bayesian log-evidence difference of $\ln B = 1.1$. The difference between these results and those obtained when including the MIRI spectrum may therefore arise not only from the additional wavelength coverage provided by MIRI, but also from differences in the spectral resolution and the treatment of inter-instrument offsets. This further highlights the sensitivity of the inferred aerosol properties to the details of the spectral data combination and the resulting degeneracies in current haze analyses. Nevertheless, in the present case, the data do not provide significant evidence in favor of a haze model, and the inferred haze properties should therefore not be interpreted as a robust detection.

\section{TOI-270~d: sensitivity to the MIRI spectrum}
\label{ap:toi270d}

TOI-270~d provides a useful test of the sensitivity of the inferred haze properties to the wavelength coverage and relative calibration of the JWST datasets. Previous analyses of the NIRISS SOSS and NIRSpec G395H observations reported detections of CH$_4$, CO$_2$, CS$_2$, and H$_2$O \citep{Benneke2024, Constantinou2026}. In our combined retrievals including the MIRI data, we recover well-constrained CH$_4$, CO$_2$, CO, and H$_2$O in the haze-free case. CO replaces CS$_2$ in our analysis because CS$_2$ is not included among the opacity sources considered in our retrieval setup.

\begin{figure}[h!]
\centering
\includegraphics[width=\hsize]{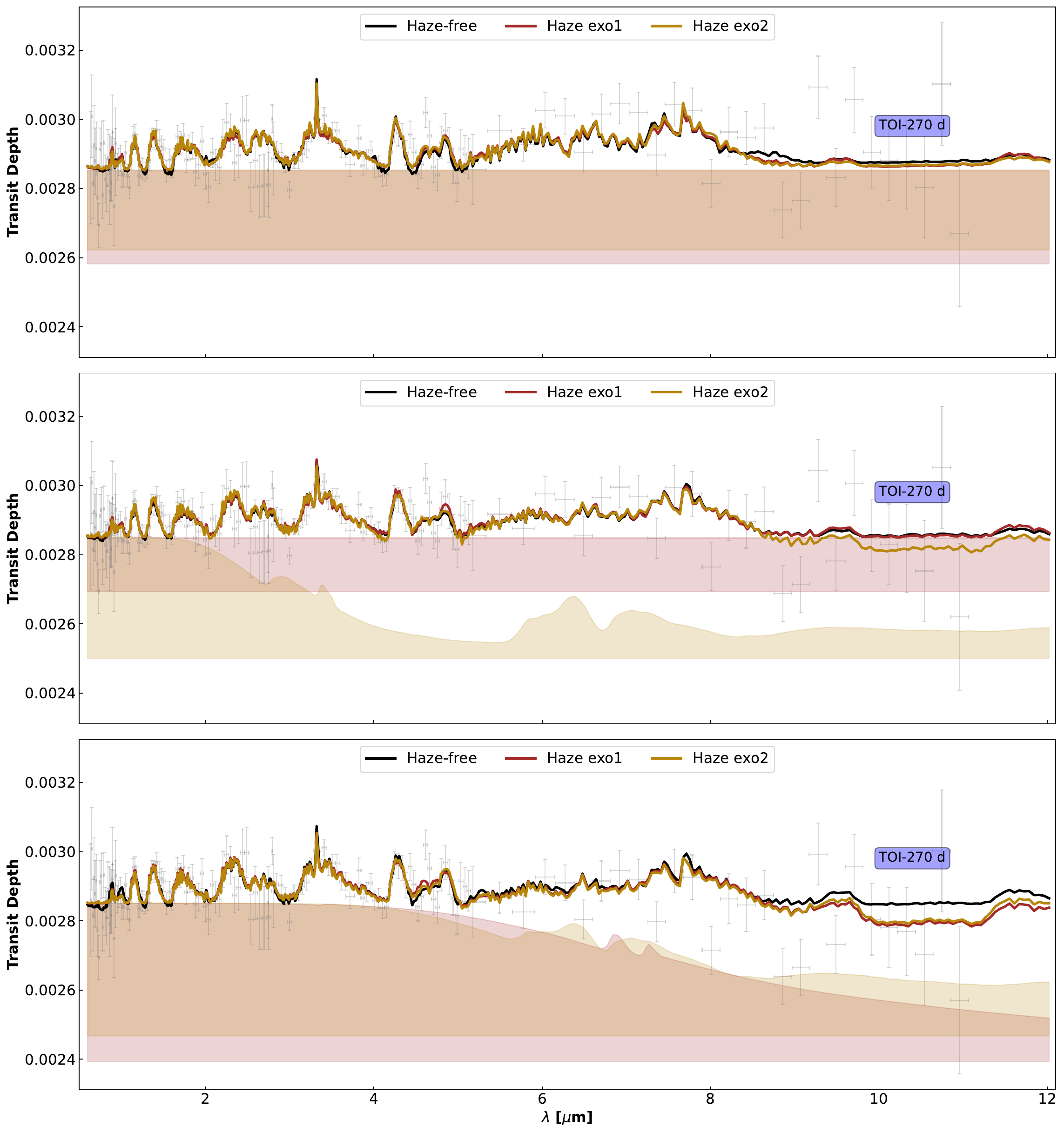}
  \caption{Observed spectra of TOI-270~d (MIRI offset; top: +50ppm, middle: 0ppm, bottom: -50ppm) shown alongside are the best-fitting models (solid lines) for haze-free (black), haze exo1 (brown) and haze exo2 (yellow) retrievals, as well as their haze contribution.}
     \label{fig:toi_miri}
\end{figure}

When haze is included, OCS becomes more constrained, but neither the exo1 nor exo2 model is statistically preferred over the haze-free retrieval for the MIRI offsets investigated. Nevertheless, the retrieved haze properties vary substantially depending on the adopted MIRI offset. For a $+50$~ppm offset, the haze layer is pushed toward low pressures and produces an approximately uniform contribution to the transmission spectrum, effectively flattening the short-wavelength spectrum. A similar behavior is obtained for the other offsets, although the magnitude of the haze contribution varies, with the largest contribution obtained when no offset is applied to the MIRI data (Fig.~\ref{fig:toi_miri}).

Additional retrievals using only the NIRISS SOSS and NIRSpec G395H spectra are presented in Appendix~\ref{ap:toi_nomiri}. These retrievals further illustrate the sensitivity of the inferred haze parameters to the adopted spectral dataset and resolution. However, the haze models remain statistically disfavored, indicating that the different haze solutions obtained with and without MIRI primarily reflect the flexibility of the retrieval rather than a robust detection of aerosol opacity.

Overall, TOI-270~d therefore demonstrates that the inclusion of MIRI data and the choice of inter-instrument offset can modify the preferred haze solution without necessarily changing the statistical conclusion regarding the presence of haze. In contrast to the stronger haze solutions found for LP~791-18~c and, more tentatively, GJ~3090~b, the TOI-270~d data do not require an aerosol component. This comparison further suggests that, within our sample, the short-wavelength spectral slope is a more important driver of the haze detections than the infrared spectral features associated with the experimentally derived haze optical properties.

\end{appendix}
\end{document}